# Survival Analysis Using a 5-Step Stratified Testing and Amalgamation Routine (5-STAR) in Randomized Clinical Trials

Devan V. Mehrotra* and Rachel Marceau West

Biostatistics and Research Decision Sciences, Merck & Co., Inc., North Wales, PA, USA

* E-mail: devan_mehrotra@merck.com

ABSTRACT

Randomized clinical trials are often designed to assess whether a test treatment prolongs survival relative to a control treatment. Increased patient heterogeneity, while desirable for generalizability of results, can weaken the ability of common statistical approaches to detect treatment differences, potentially hampering the regulatory approval of safe and efficacious therapies. A novel solution to this problem is proposed. A list of baseline covariates that have the potential to be prognostic for survival under either treatment is pre-specified in the analysis plan. At the analysis stage, using all observed survival times but blinded to patient-level treatment assignment, 'noise' covariates are removed with elastic net Cox regression. The shortened covariate list is used by a conditional inference tree algorithm to segment the heterogeneous trial population into subpopulations of prognostically homogeneous patients (risk strata). After patient-level treatment unblinding, a treatment comparison is done within each formed risk stratum and stratum-level results are combined for overall statistical inference. The impressive power-boosting performance of our proposed 5-step stratified testing and amalgamation routine (5-STAR), relative to that of the logrank test and other common approaches that do not leverage inherently structured patient heterogeneity, is illustrated using a hypothetical and two real datasets along with simulation results. Furthermore, the importance of reporting stratum-level comparative treatment effects (time ratios from accelerated failure time model fits in conjunction with model averaging and, as needed, hazard ratios from Cox proportional hazard model fits) is highlighted as a potential enabler of personalized medicine. A *fiveSTAR* R package is available at https://github.com/rmarceauwest/fiveSTAR.

## 1. INTRODUCTION

Consider a typical randomized clinical trial designed to compare the effect of treatment A (test) versus treatment B (control) on a time-to-event endpoint, the latter hereafter generically referred to as survival time. We let the random variable $T_j$ denote survival time under treatment $j$ so that $S_j(t) = Pr(T_j > t)$ is the true proportion of patients in the overall target population with survival time greater than $t$ for treatment $j$. The usual null hypothesis of interest is that the survival time distributions for the two treatments are the same, typically expressed as

$$H_0: S_A(t) = S_B(t) \text{ for all } t. \tag{1}$$

With $h_j(t)$ representing the true hazard function for treatment $j$, it is easy to see that $\int_0^t h_A(u)du / \int_0^t h_B(u)du = -logS_A(t)/-logS_B(t) \equiv \theta(t)$. If $h_A(t)$ and $h_B(t)$ are proportional to each other, $\theta(t) \equiv \theta$ is the time-invariant hazard ratio and the null hypothesis can be written as $H_0: \theta = 1$. The logrank test[1-2] or, equivalently, the score test from the Cox proportional hazards (PH) model[3] with only a treatment arm indicator, remains a popular option for testing $H_0$ in randomized clinical trials.[4-7] Two well-known issues emerge when the PH assumption is non-trivially violated, as often seen[8-13] in practice: the power of the logrank test can be substantially diminished and the hazard ratio estimate from the Cox PH model can be hard to interpret.

One way to guard against a potential power loss associated with the logrank test under non-PH conditions is through the use of a weighted logrank test, commonly selected from the $G^{\rho,\gamma}$ class of tests proposed by Harrington and Fleming[14] with weight function $w(t) = \{\hat{S}(t)\}^\rho \{1 - \hat{S}(t)\}^\gamma$ for $\rho, \gamma \geq 0$, where $\hat{S}(t)$ is the Kaplan-Meier estimator of the pooled survival function at time t. Note that $G^{0,0} \equiv Z1$ is the logrank statistic which assigns equal weight to each event, while the statistics $G^{1,0} \equiv Z2$, $G^{1,1} \equiv Z3$ and $G^{0,1} \equiv Z4$ place more weight on early, middle and late events, respectively. If prior knowledge about the expected nature of the between-treatment difference in the overall survival functions is available at the trial design stage, then a prudent choice between these four Z statistics, or selection of an alternate $G^{\rho,\gamma}$ statistic, can be made and pre-specified in the analysis plan. However, because such prior information is usually not available, versatile combinations of weighted logrank tests into single overall tests have been proposed by several authors,[15-21] including the MaxCombo test[21] which uses the minimum of Z1, Z2, Z3 and Z4 as the test statistic when a one-tailed hypothesis test is planned and negative Zi's indicate directional support for the test versus the control treatment.

Another way to mitigate the risk of reduced power associated with use of the logrank test under non-PH conditions, and to simultaneously avoid corresponding interpretational challenges with a single hazard ratio estimate, is to test $H_0$ based on an integrated weighted average of an estimate of $S_A(t) - S_B(t)$ over a time interval $(0,\tau]$.[22] A version of this test based on equal weights at each time point, typically referred to as the restricted mean survival time (RMST) test, has been popularized within the past decade by several authors.[23-26] Note that $\psi(\tau) = \int_0^\tau [S_A(t) - S_B(t)]dt$ is the true between-treatment difference in mean survival time restricted to the first $\tau$ time units of follow-up after treatment initiation. In practice, $\psi(\tau)$ is commonly estimated non-parametrically as the difference between the area under the Kaplan-Meier curve up to time $\tau$ for each treatment. An advantage of the RMST approach is that the proportional hazards assumption is not required to deliver a clinically interpretable quantification of the comparative treatment effect. However, whether and by how much the RMST test improves upon the power of the logrank test depends on the choice of $\tau$, on the censoring patterns for the treatment arms being compared, and on the extent and nature of the departure from proportional hazards.

Combinations of weighted logrank tests and the RMST test are indeed rational alternatives to the logrank test when a departure from PH is anticipated. However, all three approaches share the following two shortcomings. First, they do not leverage a ubiquitous feature of randomized clinical trial populations, namely 'structured' patient heterogeneity (described below), and this can contribute to suboptimal power for testing $H_0$. Second, even if the data support rejection of $H_0$, none of the aforementioned approaches are designed to readily deliver quantitative metrics that aid in the assessment of whether the test treatment is likely to be survival-prolonging for all types of patients in the target population or only an identifiable subset.

The above observations motivate development of an alternate approach for survival analysis in randomized clinical trials. To head in that direction, with $Y_j = log(T_j)$, we begin by noting that $\Delta = E(Y_A - Y_B)$ is a statistically unambiguous causal estimand; it represents the expected within-patient difference in log survival time between treatments A and B if each patient could hypothetically be observed under each treatment. In a parallel arm randomized clinical trial, only one of $Y_A$ and $Y_B$ can be observed for each patient, but $\Delta$ can still be estimated with minimal bias under standard assumptions since $E(Y_A - Y_B) = E(Y_A) - E(Y_B)$.

Next, we envision the overall target patient population as being a finite mixture of distinct subpopulations, which we refer to as 'risk strata'. Higher to lower ordered risk strata comprise of patients with clinical prognoses of shorter to longer expected survival regardless of assigned treatment. Patients within a given risk stratum are prognostically homogeneous in that they have in common certain pre-treatment characteristics that jointly strongly associate with survival time. In statistical parlance, the survival times under a given treatment for patients within a risk stratum are presumed to follow a common distribution. With random variable $T_{ij}$ denoting the true survival time under treatment $j$ for risk stratum $i$ ($1 \leq i \leq s$) and $Y_{ij} = log(T_{ij})$, we further envision that $Y_{iA}$ is distributed as $Y_{iB} + \Delta_i$ so that $\Delta_i = E(Y_{iA} - Y_{iB})$ represents the expected within-patient difference in log survival time under treatments A and B within stratum $i$. This implies[27] that $S_{iA}(t) = S_{iB}(\gamma_i t)$, where $S_{ij}(t) = Pr(T_{ij} > t)$ is the true proportion of patients in risk stratum $i$ with survival time greater than $t$ under treatment $j$ and $\gamma_i = e^{\Delta_i}$. We refer to $\gamma_i$ as the true *time ratio* for patients in risk stratum $i$. It represents a comparative treatment effect with a straightforward clinical interpretation; for example, $\gamma_i = 1.25$ means that patients in risk stratum $i$ are expected to have 25% longer survival under treatment A versus treatment B.

The conceptualization described above implies that the causal estimand can now be expressed as $\Delta = \sum_{i=1}^{s} f_i \Delta_i$, where $f_i$ is the proportion of patients in the entire target population that are in risk stratum $i$ ($\sum_{i=1}^{s} f_i = 1$). The null hypothesis of interest is reformulated as

$$H_0^*: \Delta_i = 0 \text{ (i.e., } \gamma_i = 1\text{) for all } i \qquad (2)$$

which is equivalent to

$$H_0^*: \cap_{i=1}^s [S_{iA}(t) = S_{iB}(t)] \text{ for all } t. \tag{3}$$

Note that if $H_0^*$ in (2) is true, then $H_0$ in (1) is also true. Furthermore, if all the $\Delta_i$'s have the same sign (an assumption that we do not require nor make), then $H_0^*$ is equivalent to

$$H_0^{**}: \Delta = 0 \text{ (i.e., } \gamma = 1), \tag{4}$$

where $\gamma = e^\Delta$ is the average (geometric mean) time ratio for the overall population.

We propose the following approach to test the null hypothesis in (2) and to estimate the causal estimand $\Delta$ (and subsequently $\gamma$). A list of baseline covariates that have the potential to be prognostic for survival under either treatment is pre-specified in the analysis plan (Step 1). At the analysis stage, using all the observed survival times but blinded to patient-level treatment assignment, 'noise' covariates are removed with elastic net Cox regression (Step 2). The shortened covariate list is subsequently used by a conditional inference tree algorithm to segment the heterogeneous trial population into subpopulations of prognostically homogeneous patients (risk strata) (Step 3). After patient-level treatment unblinding, a treatment comparison is done within each formed risk stratum (Step 4) and stratum-level results are combined for overall statistical inference (Step 5).

Before proceeding further, it is important to differentiate the proposed strategy from the common use of the stratified logrank test for hypothesis testing and corresponding stratified Cox PH model for estimation based on pre-specified stratification factor(s). While this is generally a step in the right direction relative to the unstratified version of the logrank test and Cox PH model, we caution against this routine practice for two reasons. First, inclusion of a stratification factor that does not materially influence survival time under either treatment or exclusion of a stratification factor that does so can reduce power due to over-stratification or under-stratification, respectively, both due to model misspecification.[28-29] Second, even with a proper choice of stratification factor(s), the stratified logrank test can suffer from a notable power loss either if the assumption of proportional hazards within each stratum is incorrect, or if the assumption is correct but the true hazard ratio is not constant across the strata. The primary reason for the statistical inefficiency in the latter case is that the inverse-variance weighting scheme implicitly used to combine estimated stratum-level log hazard ratios for overall inference is suboptimal when there is a treatment by stratum interaction on the hazard ratio scale.[30] In our proposal, instead of using pre-stratification and a corresponding stratified logrank/Cox PH model analysis, we use unbiased post-stratification in tandem with a new analysis approach that is intended to boost power for detecting a true between-treatment difference in the distribution of survival times. Furthermore, for the primary analysis, we quantify comparative treatment effects using time ratios instead of hazard ratios, without any need for the proportional hazards assumption anywhere, and our null hypothesis and causal estimand of interest are clearly stated with a straightforward clinical interpretation.

The rest of this article is structured as follows. In Section 2, we provide additional details for our proposed 5-step stratified testing and amalgamation routine (5-STAR). We subsequently contrast its performance with that of the unstratified logrank/Cox PH model, [pre-]stratified logrank/Cox PH model and RMST comparison strategies using a hypothetical and two real clinical trial datasets in Section 3 and simulations in Section 4. Concluding remarks are provided in Section 5.

## 2. 5-STAR DETAILS

The main concept behind our proposed 5-STAR approach is depicted schematically in Figure 1. Additional details for each step are given below.

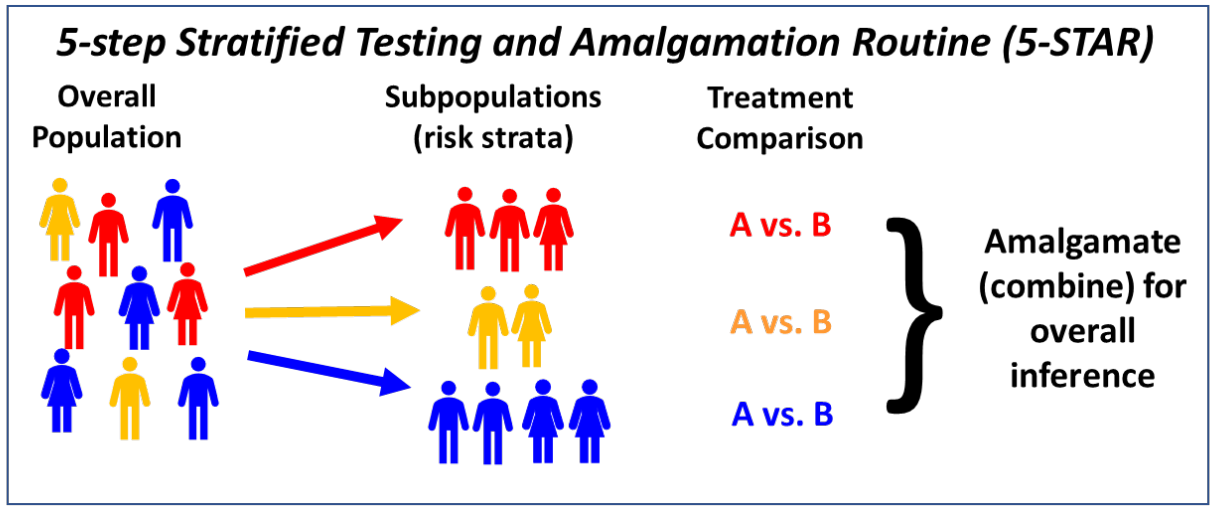

*Figure 1. 5-STAR schema*

Step 1: Pre-specify baseline covariates that may be prognostic for survival

This step entails pre-specifying in the statistical analysis plan, ideally in collaboration with relevant subject matter experts, a list of baseline covariates that have the potential to influence survival times under either treatment. The proposed methodology can easily accommodate a reasonably large number of candidate covariates of differing types (continuous, ordinal, or nominal), so more inclusivity is recommended. However, this should be done within bounds of sound statistical, clinical and operational judgement, such as avoiding the nomination of candidate covariates that are unlikely to be used in routine clinical practice due to high cost, patient inconvenience or other reasons. We assume all prognostic covariates that define the true risk strata described in the Introduction, while unknown, are included in the candidate set.

Step 2: Remove 'noise' covariates from the pre-specified candidate list

In this step, while still blinded to patient-level treatment assignment, covariates without sufficient evidence of an association with observed survival times are removed from the candidate list developed in Step 1. We considered two ways to implement this covariate filtering step: (i) random survival forest with variable importance measures and associated bootstrap confidence intervals[31-32] and (ii) elastic net Cox regression.[33-34] Since (i) and (ii) had similar performance during the initial stages of this research (results not shown), we chose the latter due to computational simplicity. The objective function for elastic net Cox regression is

$$\max_{\beta_1\ldots\beta_p}\left\{\frac{2}{N}logL(t_1\ldots t_N;x_1\ldots x_p,\beta_1\ldots\beta_p)-\lambda\left[\psi\sum_{k=1}^{p}|\beta_k|+\left(\frac{1-\psi}{2}\right)\sum_{k=1}^{p}\beta_k^2\right]\right\}, \quad (5)$$

where $L(\cdot)$ is the Cox partial likelihood function based on the pooled potentially censored survival times $(t_1\ldots t_N)$ and pre-specified baseline covariates $(x_1\ldots x_p)$, $\lambda\geq 0$ is a tuning parameter and $\psi\in[0,1]$ is a mixing parameter. In (5), $\psi=0$ and $\psi=1$ yield objective functions for Cox regression with the well-known ridge[35] and lasso[36] penalties, respectively. The elastic net combines the strengths of the two approaches.

To identify the optimal $\psi$ and $\lambda$, we consider a grid of $\psi$ values $\{0.05, 0.1, 0.15, \ldots, 0.95\}$. For each fixed $\psi$, we perform 10-fold cross-validation to determine the optimal value of $\lambda$. Here, optimal can be defined either as the value of $\lambda$ that directly minimizes the cross-validation deviance from the Cox partial likelihood model (denoted as $\lambda_{min}$) or as the largest value of $\lambda$ which yields cross-validation error that is within one standard error of the minimum deviance (denoted as $\lambda_{1se}$); we use the former as a default for 5-STAR to reduce the risk of incorrectly eliminating prognostic covariates. Based on the optimized $\psi$ and $\lambda$, all covariates with a non-zero coefficient in the final elastic net Cox regression model fit are advanced to Step 3.

Step 3: Segregate the overall population into subpopulations (risk strata) of prognostically homogeneous patients

After covariate filtering has been performed, still blinded to patient-level treatment assignment, the conditional inference tree (CTree) tool, an unbiased recursive partitioning algorithm developed by Hothorn et al,[37] is used to segregate patients into well-separated risk strata. Briefly, for each covariate $X_j$, a permutation test is performed to assess the null hypothesis that the distribution of the pooled logrank scores conditional on $X_j$ is the same as the marginal distribution of the pooled logrank scores, resulting in p-value $p_j$. If at least one of the p-values is deemed statistically significant after a Sidak[38] multiplicity adjustment, the algorithm splits the data based on the covariate with the strongest association (i.e., with the smallest p-value). If the given covariate is binary, splitting into two nodes is straightforward. Otherwise, the split point is chosen to maximize the separation in the distribution of the logrank scores between the two forming nodes. The algorithm continues doing this separately under each formed node, switching between covariate selection and split determination. Newly formed nodes are grown in this manner until the null hypothesis of no association between the pooled logrank scores and any covariate in contention cannot be rejected or until predetermined criteria (e.g., requiring at least 40 patients per terminal node) are violated. Each terminal node represents a formed risk stratum.

The CTree algorithm is a very useful patient segmentation tool. However, given the nature of the covariate-based sample splitting, it is possible to see diverging nodes from secondary splits which have similar survival profiles, visualized as approximately overlapping Kaplan-Meier curves. This can result in potential over-stratification. To avoid this, the 5-STAR method considers two iterations of the CTree algorithm, named Step 3A and Step 3B. In Step 3A, all covariates surviving the elastic net filtering step are input into the CTree algorithm to form preliminary patient risk strata, as described above. These preliminary nodes are then sorted from highest to lowest risk based on the area under the overall Kaplan-Meier curve from time zero until the minimax survival time (i.e., the minimum of the maximum observed survival times over all treatment-pooled strata) for patients within a given preliminary risk stratum. This results in an ordinal stratum variable, with value 1 for patients in the highest risk preliminary risk stratum, value 2 for patients in the second highest risk preliminary risk stratum 2, and so on. At the start of Step 3B, all patients are considered to be part of a single final risk stratum and permutation testing is performed to determine if there is sufficient statistical evidence of an association between the ordinal stratum variable and the pooled logrank scores. As in Step 3A, this second run of CTree iteratively grows a tree, finding the best split point across the ordinal stratum variable until meeting the stopping criteria. Of note, a split can separate, for example, ordered preliminary strata (1, 2) from (3, 4) but not (1, 3) from (2, 4). If all ordered preliminary strata are deemed by the algorithm to have statistically different survival profiles, the final formed strata are the same as the preliminary strata (i.e., all possible splits occur in Step 3B). Otherwise, those that are not deemed statistically different are pooled in an order restricted manner to form a smaller number of final strata. The multiplicity-adjusted p-value thresholds for allowing nodal splits need to be pre-specified; we use $\alpha_{3A} = 0.10$ and $\alpha_{3B} = 0.20$ as a default in Steps 3A and 3B, respectively.

At the end of Step 3A and 3B, still blinded to patient-level treatment assignment, suppose a total of $c$ final risk strata are formed. A graph with overlaid Kaplan-Meier curves for each of the $c$ formed risk strata can be helpful in judging the effectiveness of this critical risk-based patient segmentation step of 5-STAR; we show this with the three examples in the next Section.

Step 4: Treatment comparison within each formed risk stratum

In this step, patient-level treatment unblinding is done to enable a treatment comparison within each of the formed risk strata. Specifically, with $Y_{qk}$ denoting the log survival time for patient $k$ within formed risk stratum $q$ ($1 \leq q \leq c$), we consider a basic log-linear form of the accelerated failure time model

$$Y_{qk} = \mu_q + \delta_q I_{qk} + \sigma_q \epsilon_{qk} \qquad (6)$$

where $I_{qk}$ is a treatment indicator (1 for treatment A and 0 for treatment B), $\epsilon_{qk}$ is a random error term with an unknown density function, and $\mu_q$ and $\sigma_q$ are intercept and scale parameters, respectively. The true comparative treatment effect for patients in formed risk stratum $q$ is $\delta_q$. Note that each $\delta_q$ parameter is conditional on the overall population being represented as a mixture of subpopulations defined by the formed risk strata. While the formed risk strata will ultimately coincide with the true risk strata (conceptualized in the Introduction) with increasing sample size, this may not be the case for a given randomized clinical trial. Fortunately, this does not interfere with our main goal of estimation and inference for $\Delta$ as long as we can obtain reliable estimates for each $\delta_q$; we return to this point when we discuss Step 5.

While non-parametric analysis options for (6) do exist,[39] our preference, motivated by support from real datasets, is to use parametric approaches. Liao and Liu[40] have reported that treatment-specific Kaplan-Meier curves formed from an overall (typically heterogenous) randomized clinical trial patient population are often well approximated by parametric survival fits that assume survival times arise from a finite mixture of Weibull distributions. This mimics our own experience with the exploration of clinical trial datasets across different therapeutic areas. Given this, it is tempting to consider only a Weibull model fit for (6) in formed stratum $q$. However, we employ a model-averaging idea to add a layer of robustness against misspecification of the assumed underlying survival time distribution. Specifically, we recommend estimating $\delta_q$ through a Weibull model (model 1), a log-normal model (model 2) and a log-logistic model (model 3), resulting in a point estimate $\hat{\delta}_{q,m}$ and corresponding variance $V_{q,m}$ associated with model $m=1,2,3$. Here, the aforementioned Weibull, log-normal and log-logistic distributions refer to the distribution of survival times, with corresponding distributions for log survival times, or more specifically, for the random error terms in (6) being the extreme value (Gumbel), normal and logistic, respectively. While other parametric survival time distributions (e.g., gamma) can also be considered for (6), our experience suggests that the aforementioned three will generally suffice. Either way, it is important to stress that the parametric distributions to be used in the model-averaging need to be pre-specified in the analysis plan.

The final point estimate of $\delta_q$ and corresponding variance are obtained using the following well-known[41] model-averaging formulas based on $M$ model fits:

$$w_{q,m} = \frac{e^{-0.5 AIC_{q,m}}}{\sum_{m=1}^{M} e^{-0.5 AIC_{q,m}}} \tag{7}$$

$$\hat{\delta}_q = \sum_{m=1}^{M} w_{q,m} \hat{\delta}_{q,m} \tag{8}$$

$$V_q = \left[\sum_{m=1}^{M} w_{q,m} \sqrt{V_{q,m} + \left(\hat{\delta}_{q,m} - \hat{\delta}_q\right)^2}\right]^2. \tag{9}$$

Above, $AIC_{q,m}$ is a measure of the goodness-of-fit for parametric model $m$, quantified using the population Akaike Information Criterion, and $w_{q,m}$ is the associated weight assigned to the fit from model $m$, all within formed risk stratum $q$. Based on large-sample theory,

$$\frac{(\hat{\delta}_q - \delta_q)}{\sqrt{V_q}} \sim N(0,1) \tag{10}$$

so that an approximate $100 \times (1 - \alpha)\%$ confidence interval for $\delta_q$ is $\hat{\delta}_q \mp Z_{\alpha/2}\sqrt{V_q}$. The point estimate and confidence interval for the corresponding time ratio (i.e., $\gamma_q = e^{\delta_q}$) are easily obtained using exponentiation. In addition to the confidence interval for $\gamma_q$, we recommend reporting an estimate of the probability that treatment A prolongs expected survival relative to treatment B in formed risk stratum $q$, i.e., $Pr(\delta_q > 0) = Pr(\gamma_q > 1)$, calculated using (10).

A natural alternative to the model in (6) is the popular Cox proportional hazards model. We prefer the former because there is no guarantee that the PH assumption will hold in each formed risk stratum, nor asymptotically in each of the true risk strata. There is, however, one exception which applies if the survival times for treatment B in formed risk stratum $q$ follow a Weibull distribution and treatment A shifts the location of the log survival times under treatment B by $\delta_q$, as in (6). In this special case, it can be shown that the following identities will hold: $log[h_{qA}(t)] = log[h_{qB}(t)] + \beta_q$ and $S_{qA}(t) = [S_{qB}(t)]^{\theta_q}$, where $\beta_q = -\delta_q/\sigma_q$ and $\theta_q = e^{\beta_q}$ is the familiar hazard ratio. Given our observation that a Weibull assumption for survival times within a prognostically homogeneous subpopulation will often (but not always) be reasonable, as a supplemental analysis, we recommend reporting a point estimate and confidence interval for each $\theta_q$ and an estimate of $Pr(\theta_q < 1)$, all calculated using basic semi-parametric Cox PH model fits within each formed risk stratum. It should be recognized that these supplemental results will be hard to interpret if the PH assumption is clearly not supported by the data in one or more of the formed risk strata.

Step 5: Amalgamation of results across formed risk strata for overall hypothesis testing and estimation

After comparative treatment effect results within each formed risk stratum have been obtained, we are now ready to test $H_0^*$ in (2) and to estimate $\Delta$ (and subsequently $\gamma$). Let $n_q$ and $N$ denote the number of patients in formed risk stratum $q$ and the total number of randomized patients in the trial, respectively. The test statistics

$$Z_I = \frac{\sum_{q=1}^{c} n_q \hat{\delta}_q}{\sqrt{\sum_{q=1}^{c} n_q^2 V_q}} \quad (11)$$

and

$$Z_{II} = \frac{\sum_{q=1}^{c} n_q Z_q}{\sqrt{\sum_{q=1}^{c} n_q^2}} \quad (12)$$

are natural choices for testing $H_0^*$, where $Z_q = \hat{\delta}_q/\sqrt{V_q}$. Asymptotically, both $Z_I$ and $Z_{II}$ are distributed as N(0,1) under $H_0^*$. Intuitively, for a one-tailed test in the direction favoring the test treatment, (11) will generally be more powerful than (12) if larger $\hat{\delta}_q$ are expected a priori to be associated with larger $V_q$, such as when lower risk patients, who may have fewer events and at an overall slower rate, experience greater relative benefit from the test treatment. Since this type of information is typically unavailable at the design stage of the trial, instead of adopting either $Z_I$ or $Z_{II}$ as a default, we recommend using

$$Z_{max} = max(Z_I, Z_{II}) \quad (13)$$

as the test statistic. The exact[42] probability density function (PDF) of $Z_{max}$ is

$$f(z_{max}) = 2\phi(z_{max})\Phi\left(\frac{1-\rho}{\sqrt{1-\rho^2}} z_{max}\right), \quad (14)$$

where $\rho$ is the true correlation between $Z_I$ and $Z_{II}$; $\phi(.)$ and $\Phi(.)$ are the density function and cumulative distribution function of the standard normal distribution, respectively. In practice, $\rho$ is typically very close to one and can be estimated remarkably well with

$$\hat{\rho} = \frac{\sum_{q=1}^{c} n_q^2 \sqrt{V_q}}{\sqrt{\sum_{q=1}^{c} n_q^2 V_q} \sqrt{\sum_{q=1}^{c} n_q^2}}. \quad (15)$$

An asymptotic one-tailed p-value (p) for testing $H_0^*$ in (2) is easily calculated using (11) to (15) and the null hypothesis is rejected if p < α/2 (= 0.025 by default). In large samples, rejection of $H_0^*$ can be interpreted as reliable statistical evidence of treatment A prolonging survival relative to treatment B, on average, for patients in at least one of the formed risk strata, and hence by extension in at least one of the true risk strata. The justification for the latter extension is based on the following two lines of reasoning. First, because the formed strata and true strata represent almost the same (if not identical) risk-based groupings of patients asymptotically. Second, a stratified analysis based on formed risk strata and a corresponding analysis based on (hypothetically known) true risk strata implicitly address the same estimand, $\Delta$, which is zero under $H_0^*$. This follows because, in a typical linear model analysis context, adjustment for a different set of baseline covariates changes the estimate but not the estimand.[43]

The 5-STAR point estimate of $\Delta$ is obtained as

$$\hat{\Delta} = \frac{\sum_{q=1}^{c} w_q \hat{\delta}_q}{\sum_{q=1}^{c} w_q} \qquad (16)$$

with asymptotic variance

$$V(\hat{\Delta}) = \frac{\sum_{q=1}^{c} w_q^2 V_q}{\left(\sum_{q=1}^{c} w_q\right)^2} \qquad (17)$$

where $w_q = n_q$ if $z_{max} = z_I$ and $w_q = n_q/\sqrt{V_q}$ if $z_{max} = z_{II}$. An asymptotic $100 \times (1-\alpha)\%$ confidence interval for $\Delta$ is calculated as $\hat{\Delta} \mp Z_{max,\alpha/2}\sqrt{V(\hat{\Delta})}$ where $Z_{max,\alpha/2}$ is the relevant quantile based on the distribution in (14). Finally, point and confidence interval estimates for $\gamma$ are obtained by exponentiating their counterparts for $\Delta$ described above. Corresponding results for $\theta = e^\beta$, where $\beta = \sum_{i=1}^{S} f_i \beta_i$, can be easily obtained for a supplemental analysis, as needed, using basic Cox PH model fits with each of the formed risk strata and analogs of the formulas described above.

## 3. THREE ILLUSTRATIVE EXAMPLES

To help readers check their understanding of the 5-STAR method, and to illustrate its utility, we walk through its application in great detail using a hypothetical dataset, followed by application to two real datasets. The three examples differ primarily in terms of sample size, cardinality of the candidate set of baseline covariates in Step 1 of 5-STAR, and/or evidence of non-proportional hazards in the overall (i.e., unstratified) trial population.

Example 1 (hypothetical clinical trial)

Consider a hypothetical clinical trial in which 600 patients were randomized with equal probability to either a test treatment (A) or a control treatment (B), without pre-stratification on any factor. Assuming proportional hazards, suppose that a clinically meaningful hazard ratio of 0.80 or less, loosely interpreted by some as 'at least a 20% risk reduction', was hoped for. Based on a simulated dataset, Figure 2 (left panel) shows Kaplan-Meier curves for each treatment along with a logrank test one-tailed p-value of 0.105 and estimated hazard ratio (95% CI) of 0.87 (0.70, 1.08) obtained from a basic Cox PH model. These results look both statistically and clinically disappointing. However, a crossing of the smoothed estimates of the log hazard functions and a p-value of 0.015 from the popular Grambsch and Therneau [GT] test[44] (right panel) provide cautionary evidence of non-proportional hazards. Unfortunately, even if non-PH was anticipated a priori in this hypothetical setting and one of the two alternatives to the logrank test discussed in the Introduction had been pre-specified in the analysis plan, the result would still have been disappointing: one-tailed p-values of 0.057 and 0.182 for the MaxCombo and RMST test, respectively. In essence, application of the common approaches in current practice would have resulted in a 'negative' trial due to lack of statistical evidence of an overall survival benefit of the test versus the control treatment. We now show how the application of 5-STAR here leads to a different and more appropriate conclusion.

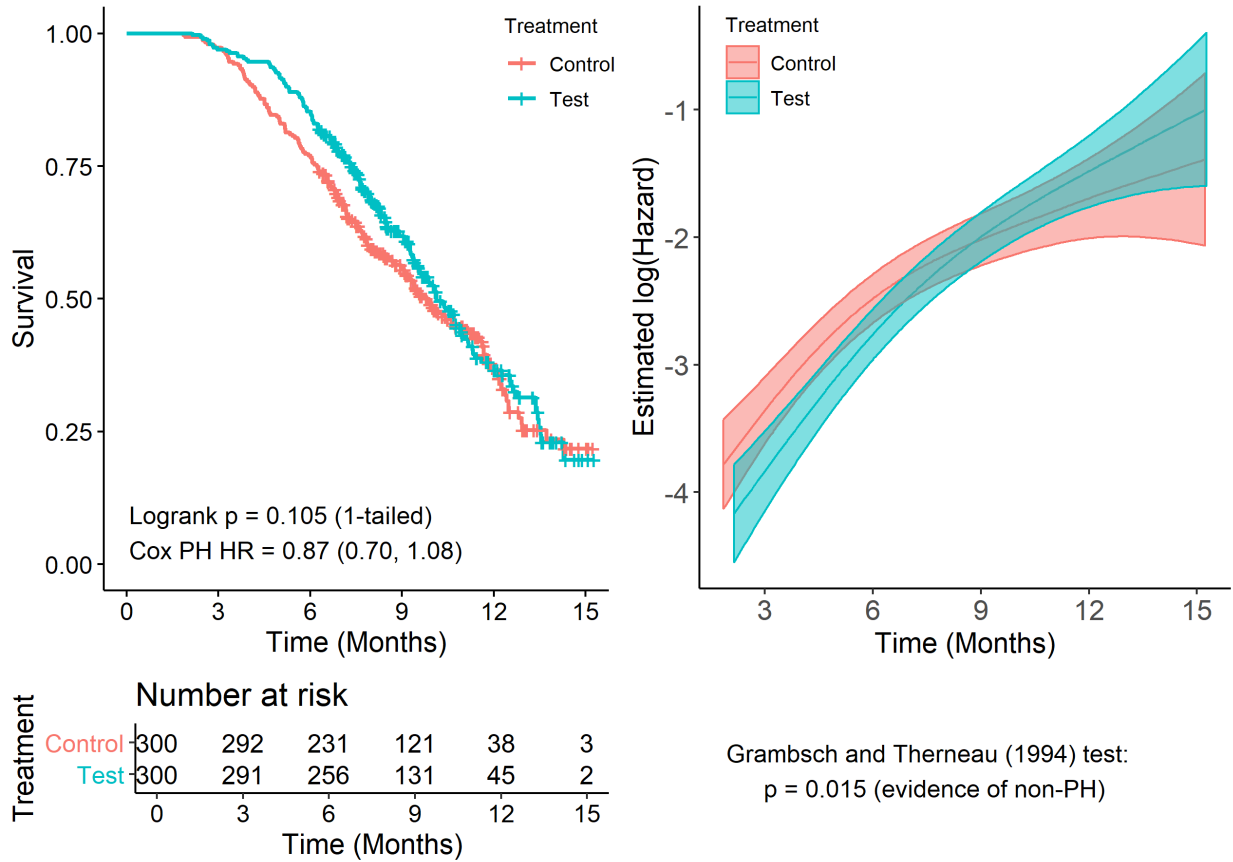

*Figure 2. Kaplan-Meier curves for each treatment (left) and corresponding smoothed versions of estimated log hazard functions (right) for Example 1.*

Suppose 50 baseline covariates, generically labeled $X_1$-$X_{50}$, were pre-specified in Step 1 and included in the hypothetical dataset. Of these, based on elastic net Cox regression applied to the pooled survival times, i.e., blinded to patient-level treatment assignment, 6 binary covariates ($X_1$, $X_2$, $X_6$, $X_7$, $X_8$, $X_{15}$) and 4 continuous covariates ($X_{26}$, $X_{31}$, $X_{38}$, and $X_{40}$) are deemed potentially prognostic for survival in Step 2. For completeness, salient details of the optimization of the elastic net mixing and tuning parameters, $(\psi, \lambda)$, and resulting 'solution path' are shown graphically in the left and right panels of Figure 3, respectively.

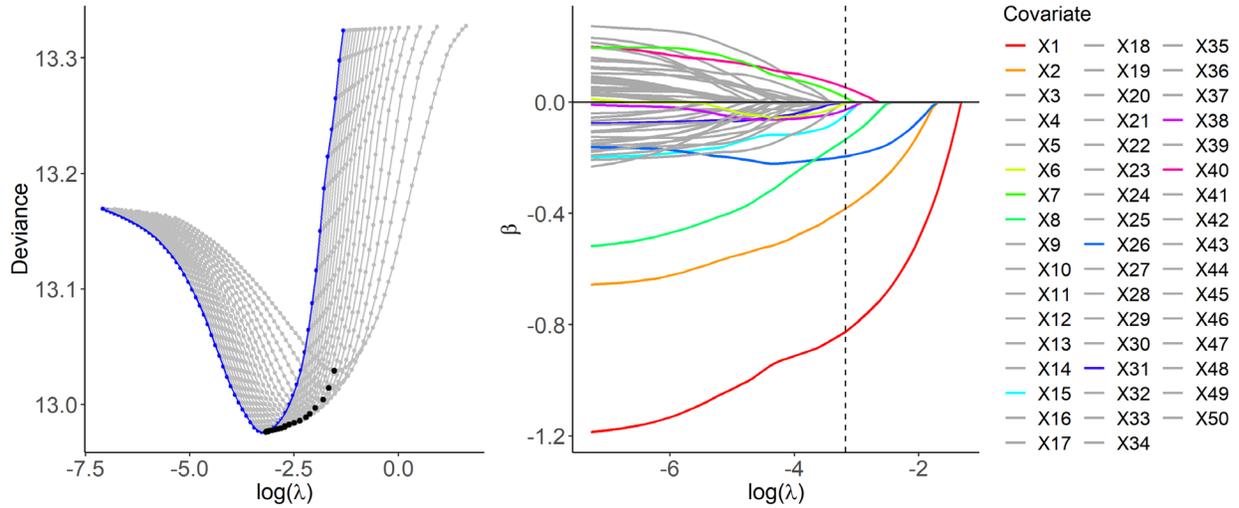

*Figure 3. Graphical details for the elastic net Cox regression filtering step (Example 1). Left: each curve represents a mixing parameter ψ and each dot represents the tuning parameter value λ with the smallest deviance. Here* $(\psi, \lambda) = (0.95, 0.04)$ *is chosen to minimize deviance. Right: solution path for the chosen α value, showing how the coefficients for the elastic net model fit change over the range of λ. Of the 50 input covariates, ten (X1, X2, X6, X7, X8, X15, X26, X31, X38 and X40) have a non-zero coefficient at the optimal λ value (shown via dotted vertical line)*

Based on the CTree algorithm results, Figure 4 reveals the formation of six preliminary (top panel; Step 3A) and four final (bottom panel; Step 3B) risk strata defined by three baseline covariates: $X_1$ (binary), $X_2$ (binary) and $X_{26}$ (continuous, with a 0.35 cut-point). The overall Kaplan-Meier curves pooled across treatment arms for the four final risk strata are clearly well-separated, with median survival times steadily increasing when transitioning in order from the highest risk stratum (S1) to the lowest risk stratum (S4).

Having reached Step 4 in the 5-STAR method, patient-level treatment unblinding can now be done. The resulting Kaplan-Meier curves for both treatments are shown within each formed risk stratum in the top panel of Figure 5. The bottom panel displays point estimates and 95% CIs for stratum-level time ratios, along with corresponding estimates of the probability that the time ratio is greater than one. All these quantities were calculated using expressions (6) through (10).

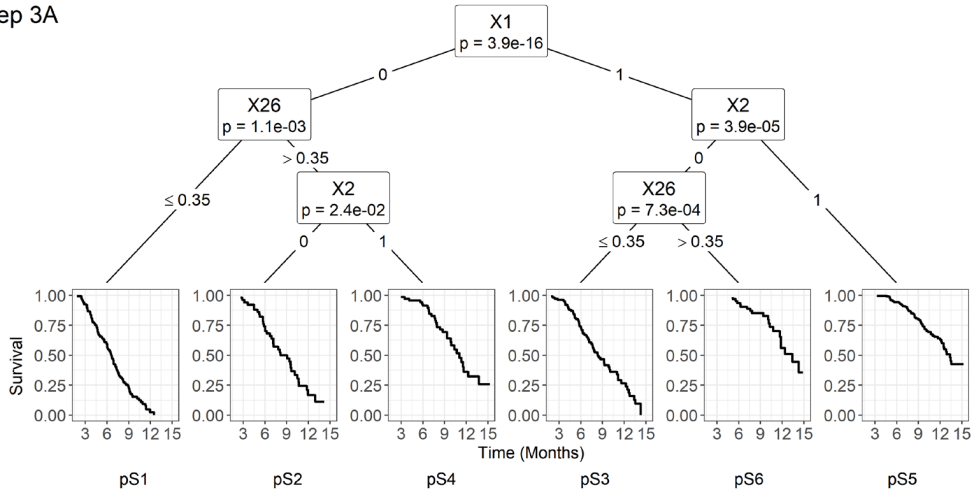
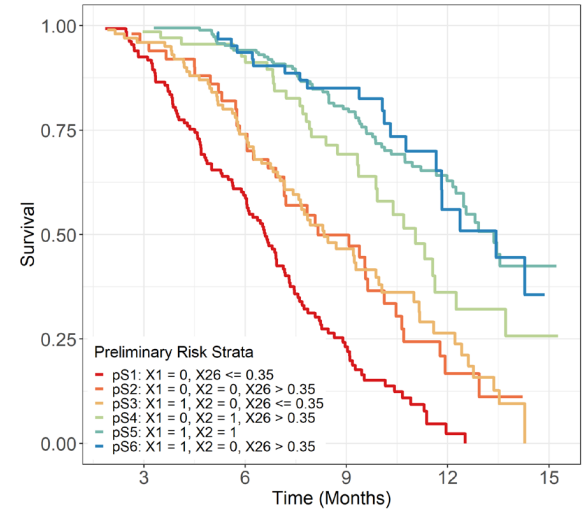
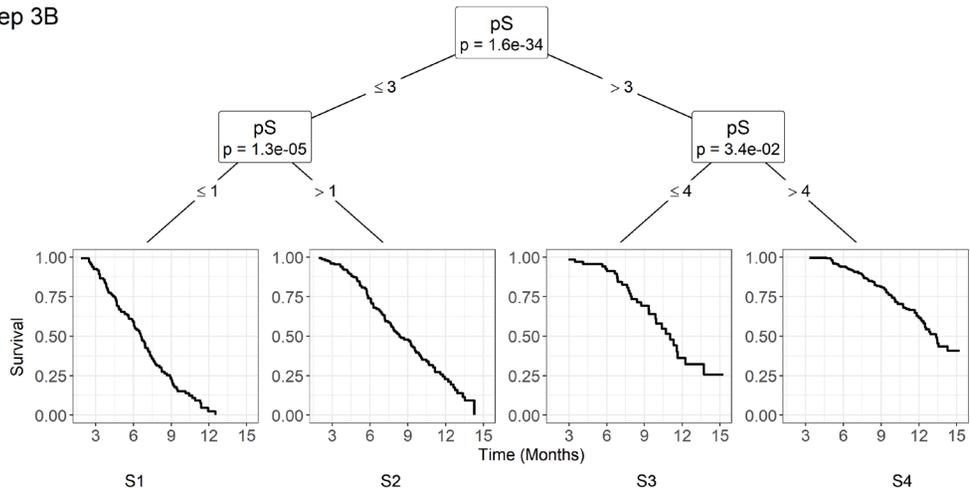
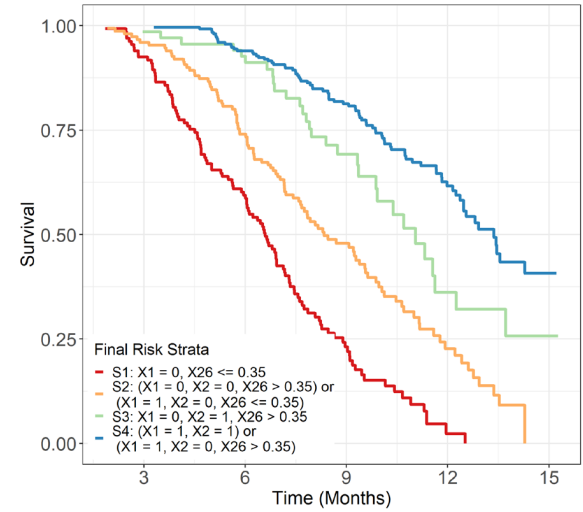

*Figure 4. Risk strata formation for Example 1 using the conditional inference tree algorithm blinded to patient-level treatment assignment. Left: results from Step 3A (preliminary risk strata) and Step 3B (final risk strata). Right: corresponding Kaplan-Meier curves for each risk stratum.*

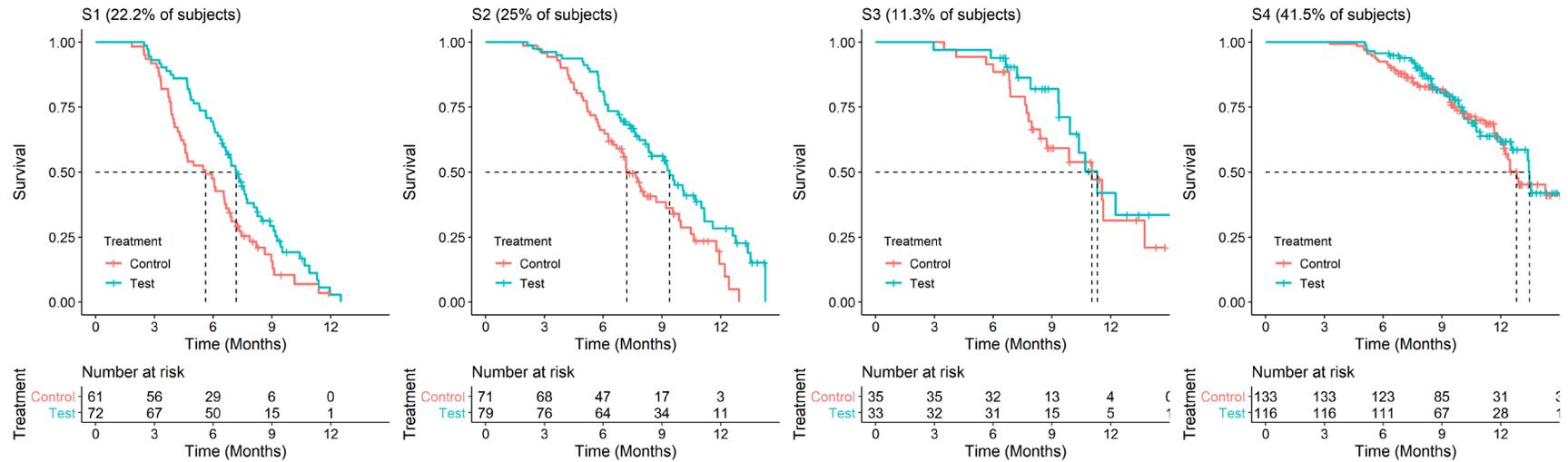

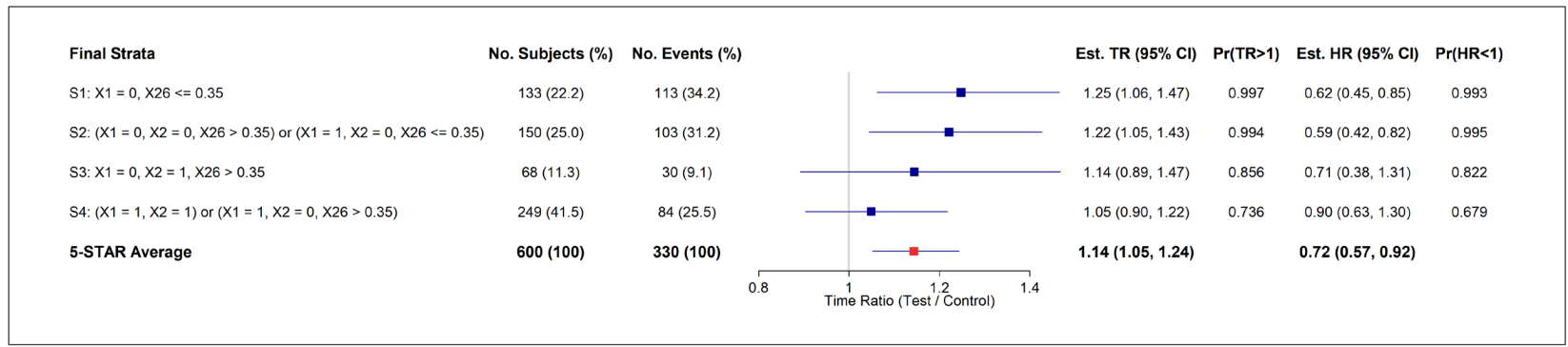

*Figure 5. Results from 5-STAR Step 4 and Step 5 for Example 1. Top: Kaplan-Meier curves by treatment within each formed risk stratum. Bottom: forest plot showing stratum-level results for each formed risk stratum as well as the overall (i.e., stratum-averaged) result. TR = time ratio, with TR > 1 favoring the test treatment. HR = hazard ratio, with HR < 1 favoring the test treatment.*

Finally, in Step 5 of 5-STAR, the individual formed risk stratum results are combined using expressions (11) through (17) leading to the estimated average time ratio (95% CI) of 1.14 (1.05, 1.24) shown in the bottom panel of Figure 5. Based on $Z_I = 3.05$, $Z_{II} = 2.95$ and $\hat{\rho} = 0.992$, the one-tailed p-value for 5-STAR is 0.001, providing strong statistical evidence that the test treatment prolongs survival relative to the control treatment, on average. This conclusion is different from that reached by the logrank, MaxCombo and RMST comparison tests.

Importantly, while 5-STAR has detected an efficacy signal favoring the test treatment over the control treatment in an 'overall' sense, it does not automatically imply that the test treatment should be deemed a better efficacy option for all types of patients. We feel it is necessary to look for subgroup(s) of patients defined by the formed risk strata that may have a 'concerningly' small likelihood of longer survival for the test treatment versus the control treatment; among several subjective options, we recommend using $Pr(TR > 1)$ less than 20% as a flagging mechanism. Based on the results shown in Figure 5, since none of the formed risk strata get flagged, it seems reasonable to conclude that the test treatment is a generally better option than the control treatment for longer survival, on average, for patients across all risk strata in the target population.

As a supplemental analysis, within each formed risk stratum, a point estimate and confidence interval for the hazard ratio along with a corresponding estimate of $Pr(HR < 1)$ are shown in the bottom panel of Figure 5. Also reported is the estimated average HR (95% CI) of 0.72 (0.57, 0.92); note that the upper bound of this CI is less than one and the point estimate is better than the aforementioned clinically meaningful threshold of 0.80. Importantly, these additional results are interpretable because no obvious departure from PH is detected within any formed risk stratum (GT p-values: 0.127, 0.987, 0.537, 0.507). Furthermore, it is reassuring to see that conclusions based on time ratios and hazard ratios are reasonably well aligned.

Example 2 (real clinical trial; oncology therapeutic area)

We now consider a real Phase III randomized clinical trial described in Lipkovich et al.[45] A total of 599 patients with a hematological malignancy were enrolled, with 303 receiving the experimental treatment and the remaining 296 receiving the control treatment. As no stratification factors were pre-specified for randomization or analysis, the [unstratified] logrank test was used for the primary analysis. This analysis missed the one-sided significance threshold of 0.025 with an overall one-tailed p-value of 0.035. The corresponding hazard ratio (95% CI), as estimated via an unstratified Cox PH model, was 0.85 (0.71, 1.01). Upon examining the Kaplan-Meier curves and estimated smoothed log hazards for each treatment (left and middle panels of Figure 6, respectively), we see hints of non-proportional hazards, with a separation in the curves occurring around the middle of the time points and crossing log hazard functions. The GT test yields a p-value of 0.002, providing statistical evidence against proportional hazards. Unlike the hypothetical example, here the MaxCombo and RMST approaches do generate a retrospective statistical 'win' for the trial with one-tailed p-values of 0.004 and 0.014, respectively. However, they are not designed to deliver a reliable and interpretable estimate of the comparative treatment effect size, either overall or in relevant subgroups of patients.

We now illustrate the application of 5-STAR to these data. In the study, 14 baseline covariates were recorded, including both nominal covariates (patient sex, race, prior therapy outcome and presence/absence of nine cytogenetic markers) and ordinal covariates (cytogenetic category and IPSS-R prognostic risk score). All these are included in Step 1. In Step 2, seven cytogenetic markers get filtered out using the elastic net Cox regression model; the remaining seven covariates are passed into the CTree algorithm, showing evidence of a possible prognostic association with survival. In Step 3A, five preliminary strata are formed. No pooling is done in Step 3B, leaving the same five final risk strata. The formed risk strata are defined using the following covariates: cytogenetic marker 6 (Cytogen6, Present (1) vs. Absent (0)), outcome of prior therapy (Priorout, with split between higher risk category "Progress" and lower risk categories {"Failure", "Relapse"}), IPSS-R prognostic score (IPSS, with split between higher risk score groups {High (3), Very High (4)} and lower risk score groups {Low (1), Intermediate (2)}), and cytogenetic category (Cytogencat, with split between lower risk group {Very Good (1), Good (2), Intermediate (3)} and higher risk group {Poor (4), Very Poor (5)}). The Kaplan-Meier curves for the five final risk strata are shown in Figure 6 (right panel). While visually there may appear to be some overlap within two pairs of curves (medium risk strata S2-S3 and low risk strata S4-S5), there is a clear separation between the three groups of curves, a corresponding clear increase in median survival, and strong separation at early time points between the aforementioned curve pairs.

After the final risk strata are formed, the within-stratum treatment effects are estimated in Step 4. The treatment-specific Kaplan-Meier curves for each formed stratum (top panel) and corresponding forest plot of estimated stratum-level time ratios and 95% confidence intervals (bottom panel) are provided in Figure 7. At the end of Step 5, the estimated average time ratio (95% CI) is 1.23 (1.08, 1.40). Furthermore, based on $Z_I = 3.15$, $Z_{II} = 2.74$ and $\hat{\rho} = 0.990$, the corresponding one-tailed p-value is 0.001. This shows strong evidence that the test treatment is beneficial in extending survival time, on average, i.e., for at least a subgroup of patients, relative to the control treatment. As in the previous example, we now take a closer look at the formed risk stratum-level results. Here, for the higher risk patients (S1-S3, corresponding to about 50% of the patients) the test treatment is estimated to prolong expected survival by an impressive 50-85% over the control treatment, with greater than 99% probability that the true time ratio exceeds 1. In contrast, there is no evidence that the test treatment will be more beneficial for the lowest risk patients (S5) where $Pr(TR > 1)$ is < 5%. In a regulatory setting, it would seem reasonable to question the approvability of the test treatment for this 12% subset of the target patient population, corresponding to patients with lower IPSS risk scores and better cytogenetic categories.

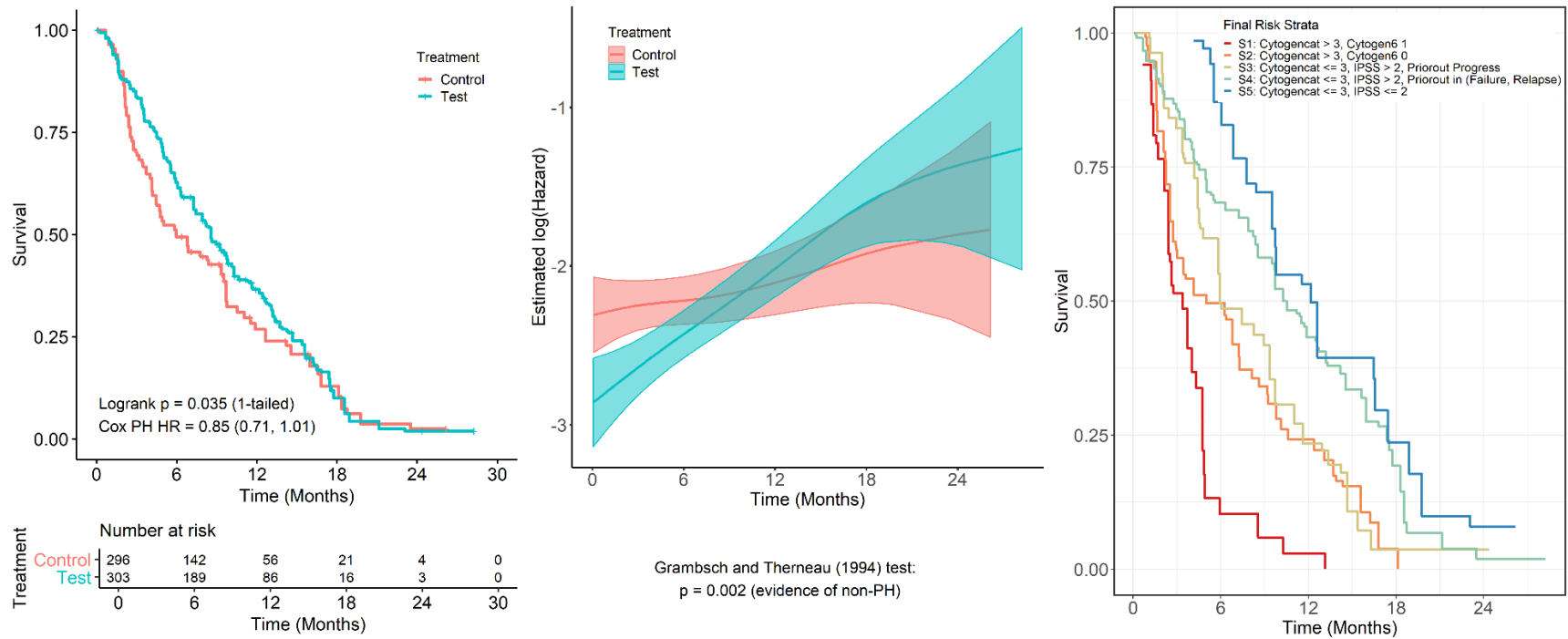

*Figure 6. Kaplan-Meier curves for each treatment (left) and corresponding smoothed versions of estimated log hazard functions (middle) for Example 2. Right: KM curves for final risk strata formed using the conditional inference tree algorithm blinded to patient-level treatment assignment.*

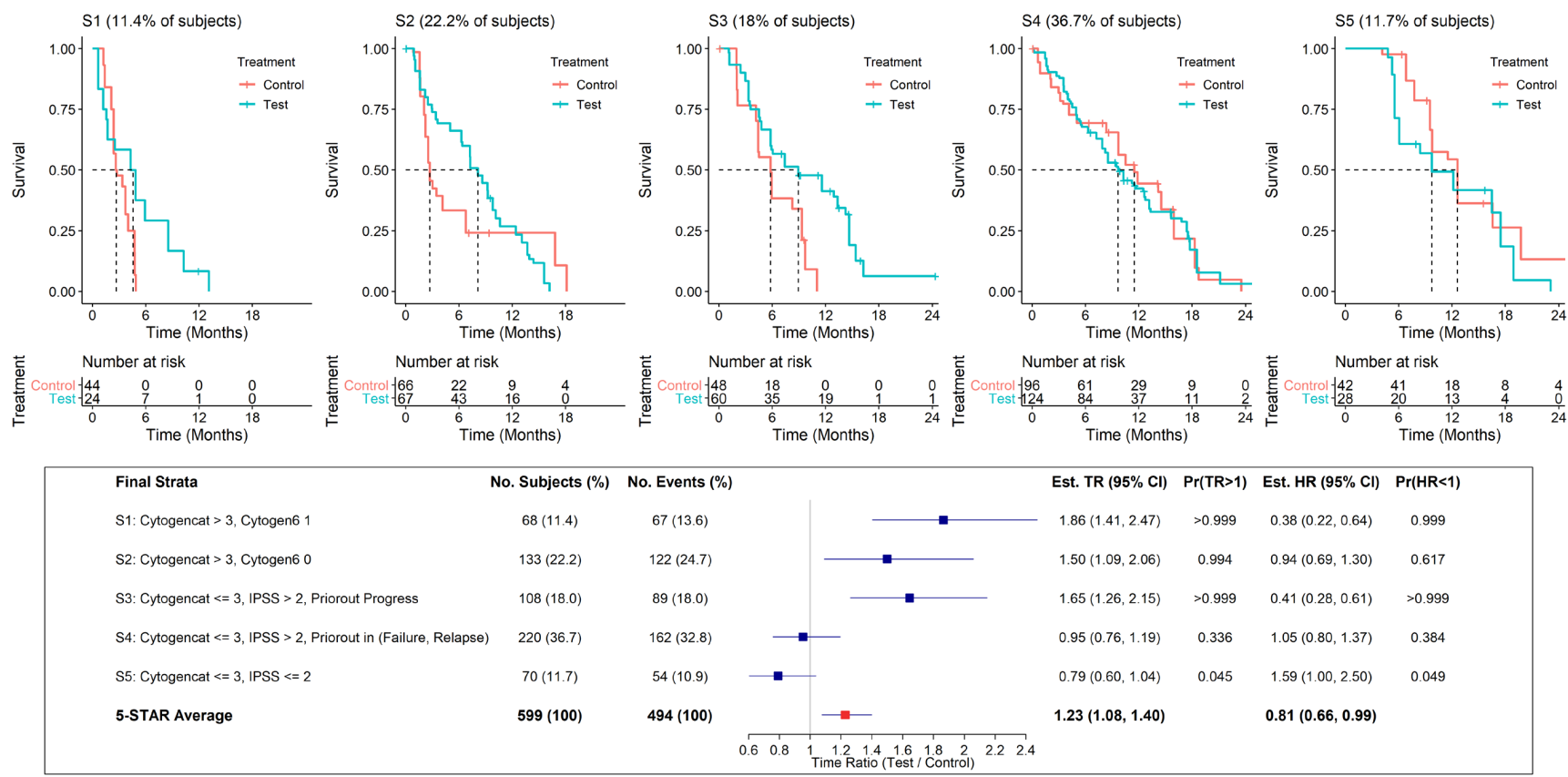

*Figure 7. Results from 5-STAR Step 4 and Step 5 for Example 2. Top: Kaplan-Meier curves by treatment within each formed risk stratum. Bottom: forest plot showing stratum-level results for each formed risk stratum as well as the overall (i.e., stratum-averaged) result. TR = time ratio, with TR > 1 favoring the test treatment. HR = hazard ratio, with HR < 1 favoring the test treatment.*

For the most part, conclusions are consistent when examining the supplemental hazard ratio results. The overall estimated average HR (95% CI) is 0.81 (0.66, 0.99), which is more promising both statistically and clinically compared to the corresponding result from the prespecified analysis (Cox PH model). Of note is the second highest formed risk stratum S2, where the estimated hazard ratio (95% CI) is 0.94 (0.69, 1.30) with less than 62% probability that the true hazard ratio is less than one. This shows materially lower evidence of test treatment versus control treatment benefit as compared to that based on the time ratio-based estimate. However, this stratum, along with strata S1 and S3 show evidence of non-PH (GT p-value < 0.02 in all three strata, and < 0.0001 in S2); this makes the hazard ratio results hard to interpret, amplifying the utility of the corresponding primary analysis results using time ratios.

Example 3 (real clinical trial; cardiovascular therapeutic area)

As a final example, we consider a large phase III cardiovascular clinical trial.[46] This randomized study enrolled 18,144 patients who had recently experienced an acute coronary syndrome: 9,067 received the test treatment and 9,077 received the control treatment (standard of care). Randomization was stratified based on three pre-defined baseline factors: PLL (prior use of lipid lowering therapy, a binary variable indicating whether or not a patient had used statins before trial initiation), ACSD (type of acute coronary syndrome experienced, STEMI (ST-Elevation Myocardial Infarction) or non-STEMI), and EACS (representing patient participation status in a previous cardiovascular trial: unenrolled, enrolled receiving test treatment, or enrolled receiving control treatment). As only non-STEMI patients were enrolled in the EACS trial, eight pre-defined strata were used for randomization and subsequent protocol-defined analyses.

For illustrative purposes, we focus on the time to stroke (fatal or non-fatal), a clinically important exploratory endpoint for the trial with over 95% censoring. The Kaplan-Meier curves for each treatment arm are shown in the left panel of Figure 8; unlike the previous two examples, here there is no evidence of non-proportional hazards, with a GT p-value of 0.479. In the middle panel, we provide the overall (i.e., pooled across treatment arms) Kaplan-Meier curves for each of the eight pre-defined strata. There is a substantial amount of overlap in some of the curves, indicating a potentially suboptimal choice of pre-stratification factor(s) leading to over-stratification. The pre-specified stratified logrank test just falls short of statistical significance at the one-sided 0.025 alpha level, with a p-value of 0.026 and corresponding HR (95% CI) of 0.86 (0.73, 1.00) as estimated via the stratified Cox PH model. The corresponding one-tailed p-values for the RMST comparison and MaxCombo test are 0.024 and 0.039, respectively.

We now apply the 5-STAR method to this large dataset. Forty-five potentially prognostic variables are specified in Step 1, including demographic, clinical disease history, and baseline lipid level information. Of these, blinded to patient-level treatment assignment, 22 pass the filtering step in Step 2 as having some evidence of association with survival time. Finally, in Step 3, five preliminary and four final risk strata are formed based on continuous variable age (with cut-point at 67 years) and three binary patient history variables: HSCD (history of cerebrovascular disease), HSAF (history of atrial fibrillation), and PRMI (prior myocardial infarction). Overall Kaplan-Meier curves for the four final formed risk strata are shown in the right panel of Figure 8. There is a much more distinct separation in the pooled survival curves compared to that in the plot with the curves based on the pre-specified strata (middle panel). In particular, the highest formed risk stratum has clearly worse prognosis compared to the other three, which still have close to no overlap. This example vividly illustrates how the objective risk-based patient segmentation component of 5-STAR can reveal 'structured' patient heterogeneity using an algorithm that is blinded to patient-level treatment assignment. This type of information, important both from a clinical and statistical perspective, often remains hidden when viewed through strata defined by (often subjectively chosen) pre-stratification factors at the design stage of the trial.

Kaplan-Meier curves for both treatments within each formed risk stratum (top panel) and a corresponding forest plot of the estimated comparative treatment effects (bottom panel) are shown in Figure 9, illustrating the components behind the overall comparative treatment effect estimate obtained in Step 5 of 5-STAR. The average estimated time ratio (95% CI) is 1.30 (1.04, 1.81), indicating that, on average, patients on the test treatment have 30% longer stroke-free time compared to those on the control treatment. Based on $Z_I = 2.32$, $Z_{II} = 2.36$ and $\hat{\rho} = 0.998$, the corresponding one-tailed p-value is 0.010, indicating statistical evidence of efficacy benefit from the test treatment compared to the control for at least some patients. Looking at the results within the formed risk strata, the lowest risk patients (i.e., those that are less than 67 years of age, with no prior MI and no history of cerebrovascular disease [S4]), representing just over half of all patients, have the largest apparent gain from taking the test treatment with an estimated time ratio (95% CI) of 1.50 (1.04, 1.61) and 99% probability of improved survival benefit ($\Pr(TR > 1)$) relative to the control treatment. The supplemental hazard ratio-based results are clinically consistent with the time ratio results, with average hazard ratio (95% CI) of 0.81 (0.67, 0.97), and an estimated hazard ratio (95% CI) in formed risk stratum S4 of 0.70 (0.55, 0.90) with over 99% probability that the true hazard ratio is less than one for such patients. Of note, none of the $\Pr(TR > 1)$ or $\Pr(HR < 1)$ values shown in Figure 9 are even close to being lower than 20%, our subjective threshold of concern introduced in Example 1. Accordingly, if these results were to serve as the primary basis for a regulatory action, we believe it would be reasonable to consider approvability of the test treatment for patients across all risk strata.

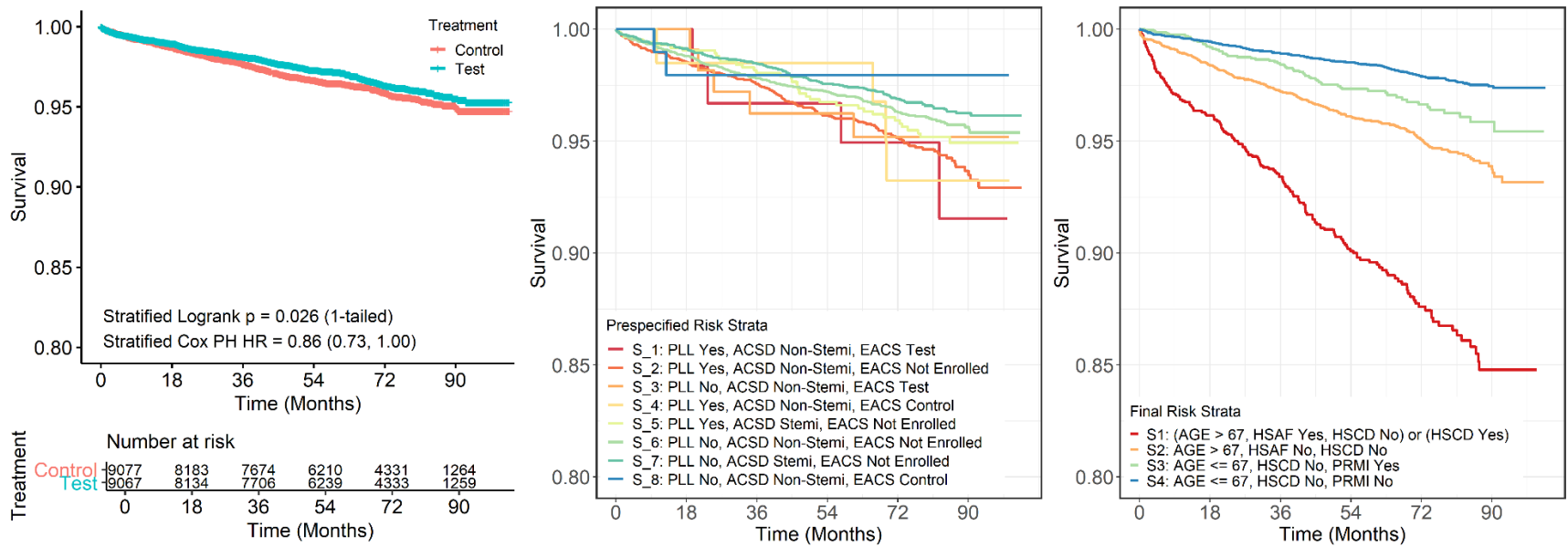

*Figure 8. Kaplan-Meier (KM) curves for Example 2. Left: KM curves for test and control treatments. Middle: KM curves for each pre-specified stratum defined in the study protocol. Right: KM curves for final risk strata formed using the conditional inference tree algorithm blinded to patient-level treatment assignment.*

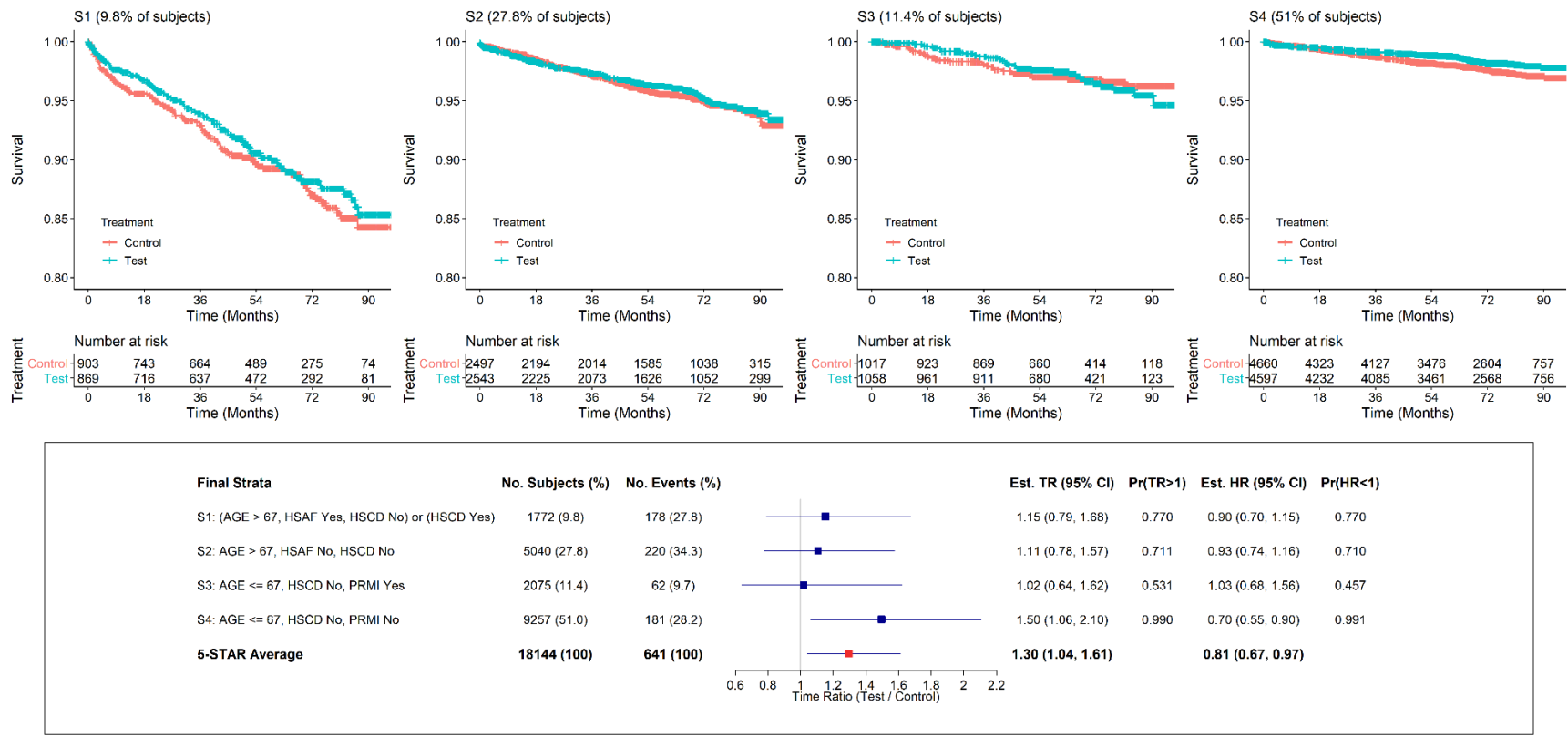

*Figure 9. Results from 5-STAR Step 4 and Step 5 for Example 3. Top: Kaplan-Meier curves by treatment within each formed risk stratum. Bottom: forest plot showing stratum-level results for each formed risk stratum as well as the overall (i.e., stratum-averaged) result. TR = time ratio, with TR > 1 favoring the test treatment. HR = hazard ratio, with HR < 1 favoring the test treatment.*

## 4. SIMULATION STUDY

### 4.1 Set-up

We simulated a clinical trial setting in which 600 patients are randomized in a 1:1 ratio to receive either a test treatment (A) or control treatment (B). For each patient, 50 correlated baseline covariates, $X_1$-$X_{25}$ binary and $X_{26}$-$X_{50}$ continuous, are measured. The heterogeneous patient population consists of four distinct prognostic subpopulations (risk strata) defined by a joint composite of two binary covariates ($X_1$ and $X_2$) and one dichotomized continuous covariate ($X_{26} \leq 0.4$), as shown in Table 1. We set the true marginal means of the binary and continuous covariates to be 0.5 and 0, respectively. Furthermore, we assumed a true pairwise correlation of 0.2 between the three aforementioned prognostic covariates and sampled from $N(0, sd = 0.15)$ for all other pairwise correlations; these correlations were motivated by real datasets across different therapeutic areas (details omitted).

**Table 1**

Simulation conditions showing four true risk strata defined by three baseline covariates, and median survival time for the control treatment, hazard ratio (HR) and time ratio (HR) within each true risk stratum

| Risk Stratum | X1 | X2 | X26 | Median Survival (control treatment) | Simulation Scenarios and Stratum-level Treatment Effects | | | | | | | |
|---|---|---|---|---|---|---|---|---|---|---|---|---|
| | | | | | Null HR = 1 | | Alt-1 Equal HRs | | Alt-2 Increasing HRs | | Alt-3 Decreasing HRs | |
| | | | | | HR | TR | HR | TR | HR | TR | HR | TR |
| S1 | 0 | 0 | ≤ 0.4 | 6.0 months | 1 | 1 | 0.7 | 1.15 | 0.42 | 1.42 | 0.95 | 1.02 |
| | 0 | 1 | ≤ 0.4 | | | | | | | | | |
| S2 | 0 | 0 | > 0.4 | 8.4 months | 1 | 1 | 0.7 | 1.13 | 0.7 | 1.13 | 0.86 | 1.05 |
| | 1 | 0 | ≤ 0.4 | | | | | | | | | |
| S3 | 0 | 1 | > 0.4 | 10.8 months | 1 | 1 | 0.7 | 1.11 | 0.86 | 1.04 | 0.7 | 1.11 |
| | 1 | 1 | ≤ 0.4 | | | | | | | | | |
| S4 | 1 | 0 | > 0.4 | 13.2 months | 1 | 1 | 0.7 | 1.09 | 0.95 | 1.01 | 0.42 | 1.24 |
| | 1 | 1 | > 0.4 | | | | | | | | | |

For each treatment group, the random number of patients enrolled from each of the four risk strata was sampled from a $multinomial(n = 300, p_1 = p_2 = p_3 = p_4 = 0.25)$ distribution. (Of note, simulations with unequal prevalence across the risk strata were also conducted and yielded similar conclusions; details omitted). A trial entry time for each patient $k = 1, \dots, n_i$ within risk stratum $i = 1, \ldots, 4$, ordered from highest risk to lowest risk, and treatment $j = A, B$ was generated as $e_{ijk} \sim Uniform(0, E)$ where $E = 0.75$ years, or 9 months. True survival times were generated from a Weibull distribution with shape and scale parameters determined by risk stratum membership and treatment assignment as $S_{ijk} \sim Weibull(shape = \kappa_i, scale = \eta_{ij})$, where $\boldsymbol{\kappa} = (\kappa_1, \kappa_2, \kappa_3, \kappa_4) = (2.5, 3, 3.5, 4)$, and $\eta_{iB}$ back-calculated from the shape and desired median survival time for the control treatment arm (in years) of each stratum, $\boldsymbol{m_B} = (m_{1B}, m_{2B}, m_{3B}, m_{4B}) = (0.5, 0.7, 0.9, 1.1)$, as $\eta_{iB} = m_{iB} \times \ln(2)^{-\frac{1}{\kappa_i}}$. Scale parameters in the test treatment arm were calculated as $\eta_{iA} = \eta_{iB} \times \theta_i^{-\frac{1}{\kappa_i}}$, where $\theta_i$ was the hazard ratio for stratum $i$. This set-up ensured proportional hazards in truth within each risk stratum.

In each simulated trial, patients were followed until 330 deaths had accrued. This target event total is often chosen in practice because it gives approximately 90% power for the logrank test when the proportional hazards assumption holds, the true hazard ratio is 0.7, and testing is done at the one-tailed alpha=0.025 level. The 'observed' simulated data for each trial were $(T_{ijk}, \delta_{ijk}, x_{ijk})$, where $T_{ijk} = \min(S_{ijk}, C - e_{ijk})$ was the observed survival time, $C$ was the random time at which 330 events had accrued, indicating end of patient follow-up, $\delta_{ijk} = I(S_{ijk} > C - e_{ijk})$ was the censoring indicator, equal to 0 for all patients whose true event time was observed and 1 for all patients who were censored, and $x_{ijk}$ was the vector of patient-level baseline covariate data.

Four scenarios were considered to evaluate the performance of the competing methods under different patterns of hazard ratios across the risk strata. First, to evaluate type I error, a null scenario was considered in which there was no treatment difference in any of the risk strata, i.e., $\boldsymbol{\theta} = (\theta_1, \theta_2, \theta_3, \theta_4) = (1,1,1,1)$. Three alternative scenarios were also considered, all with an average log hazard ratio of $\beta = \sum_{i=1}^{S} f_i \beta_i = \ln(0.7)$, with $\theta_i = e^{\beta_i}$, as noted in Section 2. In Alt-1 we set $\boldsymbol{\theta} = (0.7, 0.7, 0.7, 0.7)$, i.e., equal hazard ratios across the ordered risk strata. In Alt-2 we set $\boldsymbol{\theta} = (0.42, 0.7, 0.86, 0.95)$, i.e., increasing hazard ratios, implying larger survival benefit of the test versus the control treatment for patient subpopulations with higher risk. Finally, in Alt-3 we set $\boldsymbol{\theta} = (0.95, 0.86, 0.7, 0.42)$, i.e., decreasing hazard ratios, implying larger survival benefit of the test versus the control treatment for patient subpopulations with lower risk. All the scenarios are summarized in Table 1. In addition to the true hazard ratios for each risk stratum, we have also shown the corresponding true time ratios calculated as $\gamma_i = e^{\Delta_i}$, where $\Delta_i = -\beta_i/\kappa_i$.

We compared the performance of 5-STAR (both time ratio and hazard ratio versions) with that of the unstratified logrank test/Cox PH model, a misspecified stratified logrank test/Cox PH model, the MaxCombo and the RMST comparison methods. The misspecified stratified logrank test/Cox PH model was intended to mimic a realistic setting in which expert knowledge may guide researchers to correctly identify only a subset of the true prognostic covariates at the trial planning stage; in effect, either some prognostic covariates are missed, suboptimal cut-offs are selected for dichotomization of prognostic continuous covariates, or non-prognostic covariates are erroneously used for pre-stratification. Here, our pre-specified stratified Cox model analysis reflected strata defined by 'correct' covariates $X_2$ and $X_{26}$ (the latter discretized using a slightly suboptimal cut-off of zero instead of the truly optimal 0.4) as well as an incorrect, i.e., non-prognostic, covariate $X_3$.

In each of the four simulation scenarios, for each competing method, we computed the proportion of times in 20,000 simulated trials that the null hypothesis $H_0^*$ in (2), equivalent in this setting to $H_0^{**}$ in (4) and to $H_0$ in (1), was rejected at the one-sided 2.5% significance level incorrectly (for type I error) or correctly (for power). For 5-STAR, we also calculated mean percent bias for the average time ratio ($\gamma$) and average hazard ratio ($\theta$) estimands along with the corresponding coverage of the 95% CI for each parameter; these metrics were also calculated for the unstratified and stratified Cox PH model fits with $\theta$ serving as the target estimand. Finally, to help understand the key drivers of the main 5-STAR results, we examined the performance of the elastic net filtering step (Step 2) and the CTree step (Step 3) in terms of removal of noise covariates and use of correct (i.e., truly prognostic) covariates to define the risk strata, respectively.

**4.2 Results**

Type I error and power

As shown in Table 2, the type I error rate was well-controlled, on the basis of being less than $0.025 + 2\sqrt{(0.025 \times 0.975)/20,000} = 2.72\%$, for all the competing methods. In terms of power, both the time ratio (TR) and hazard ratio (HR) versions of 5-STAR performed admirably. Under Alt-1, both 5-STAR [TR] and 5-STAR [HR] had 84% power, which was 7 to 17 percentage points higher than the power of the best (stratified logrank) and worst (MaxCombo) performers among the other methods. Under Alt-2, 5-STAR [TR] placed first with 93% power followed by a tie for second place between 5-STAR [HR] and stratified logrank (90% power each), well-separated from the other methods which had powers ranging from 82% to 84%. Finally, under Alt-3, the best two performers were 5-STAR [HR] with 73% power and 5-STAR [TR] with 67% power, both notably more powerful than the other methods which had powers ranging from 48% to 54%.

**Table 2**

Summary of simulation results: type I error (target α=2.5%), power, and mean percent bias and coverage of 95% CI for the relevant target estimand (average HR or TR) based on 20,000 simulations under each scenario

| Scenario:<br>True HRs in risk strata: | Null<br>1, 1, 1, 1 | Alt-1<br>0.7, 0.7, 0.7, 0.7 | | | Alt-2<br>0.42, 0.7, 0.86, 0.95 | | | Alt-3<br>0.95, 0.86, 0.7, 0.42 | | |
|---|---|---|---|---|---|---|---|---|---|---|
| Analysis Method | Type I<br>Error (%) | Power<br>(%) | Mean<br>% Bias | 95% CI<br>Coverage | Power<br>(%) | Mean<br>% Bias | 95% CI<br>Coverage | Power<br>(%) | Mean<br>% Bias | 95% CI<br>Coverage |
| Unstratified logrank/CPH | 2.56 | 71 | 9.1 | 88.3 | 82 | 4.9 | 93.3 | 50 | 15.9 | 74.9 |
| Stratified logrank/CPH* | 2.49 | 77 | 6.1 | 92.3 | 90 | -0.3 | 95.0 | 48 | 15.9 | 76.1 |
| MaxCombo | 2.60 | 67 | -- | -- | 83 | -- | -- | 54 | -- | -- |
| RMST | 2.51 | 71 | -- | -- | 84 | -- | -- | 48 | -- | -- |
| 5-STAR [TR] | 2.49 | 84 | 0.3 | 95.8 | 93 | 0.2 | 95.3 | 67 | 0.3 | 96.0 |
| 5-STAR [HR] | 2.52 | 84 | 0.4 | 95.9 | 90 | -2.7 | 94.9 | 73 | 1.2 | 96.1 |

CPH = Cox proportional hazards model, * analysis strata based on three pre-specified factors (two prognostic and one non-prognostic, in truth), TR = time ratio, HR = hazard ratio

Percent bias and coverage of 95% CI for target estimand

As shown in Table 2, point estimates using 5-STAR [TR] and 5-STAR [HR] were associated with negligible bias for their respective average time ratio (γ) and average hazard ratio (θ) target estimands. Moreover, the 5-STAR 95% confidence intervals captured their true target estimand very close to 95% of the time in every scenario studied. In contrast, the unstratified and stratified Cox PH analyses were associated with point estimates for θ that were notably biased under Alt-1 and Alt-3, with corresponding 95% confidence intervals having considerable under-coverage of the target parameter.

5-STAR: effectiveness of covariate filtering (Step 2) and risk strata formation (Step 3)

The elastic net Cox regression algorithm using pooled survival times blinded to patient-level treatment assignment did a reasonably good job filtering out noise covariates in Step 2 of 5-STAR. Across the four simulated scenarios, on average, approximately 9-10 of the 50 candidate covariates advanced to Step 3, with all of the correct (i.e., truly prognostic) covariates $X_1$, $X_2$ and $X_{26}$ advancing approximately 94-99% of the time. The CTree algorithm used for stratum formation in Step 3 of 5-STAR also performed reasonably well. Across the four simulated scenarios, on average, 3.4 covariates defined the final formed risk strata, with the latter based on at least the three correct covariates 83-98% of the time and based on only the three correct covariates 48-64% of the time.

## 5. Concluding remarks

The power of the ubiquitous logrank test for a between-treatment comparison of survival times in randomized clinical trials can be notably less than desired if the treatment hazard functions are non-proportional, and the accompanying hazard ratio estimate from a Cox proportional hazards model can be hard to interpret. Increasingly popular approaches to guard against the statistical adverse effects of non-proportional hazards include the MaxCombo test (based on a versatile combination of weighted logrank statistics) and a test based on a between-treatment comparison of restricted mean survival time (RMST). Unfortunately, neither the logrank test nor the latter two approaches are designed to leverage what we refer to as structured patient heterogeneity in clinical trial populations, and this can contribute to suboptimal power for detecting a between-treatment difference in the distribution of survival times. Stratified versions of the logrank test and the corresponding Cox proportional hazards model based on pre-specified stratification factors represent steps in the right direction. However, they carry unnecessary risks associated with both a potential suboptimal choice of stratification factors and with potentially implausible dual assumptions of proportional hazards within each stratum and a constant hazard ratio across strata.

We have developed and described a novel alternative to the aforementioned current approaches for survival analysis in randomized clinical trials. Our approach envisions the overall patient population as being a finite mixture of subpopulations (risk strata), with higher to lower ordered risk strata comprised of patients having shorter to longer expected survival regardless of treatment assignment. Patients within a given risk stratum are deemed prognostically homogeneous in that they have in common certain pre-treatment characteristics that jointly strongly associate with survival time. Given this conceptualization and motivated by a reasonable expectation that detection of a true treatment difference should get easier as the patient population gets prognostically more homogeneous, our proposed method follows naturally. Starting with a pre-specified set of baseline covariates (Step 1), elastic net Cox regression (Step 2) and a subsequent conditional inference tree algorithm (Step 3) are used to segment the trial patients into ordered risk strata; importantly, both steps are blinded to patient-level treatment assignment. After unblinding, a treatment comparison is done within each formed risk stratum (Step 4) and stratum-level results are combined for overall estimation and inference (Step 5).

For the primary analysis, labeled 5-STAR [TR], exponentiated estimates of between-treatment differences in mean log survival time, referred to as time ratios, are used for treatment comparisons within the formed risk strata. Estimation is accomplished using three accelerated failure time model fits (assuming survival times follow either a Weibull, log-normal or log-logistic distribution) in conjunction with straightforward model averaging. As a supplemental analysis, labeled 5-STAR [HR], hazard ratio estimates from basic Cox proportional hazard model fits within formed risk strata can be used, with the usual understanding that a hazard ratio estimate is hard to interpret if the corresponding treatment hazard functions are non-proportional. In addition to each formed stratum-level point estimate and 95% confidence interval for a time ratio (and hazard ratio, if needed), we recommend reporting a corresponding estimate of the probability that the test treatment is associated with expected longer survival than the control treatment. This level of detail in stratum-level reporting, currently uncommon in practice, provides transparency that aids in understanding the key inputs for the overall (i.e., stratum-averaged) result. Moreover, it serves as a potential enabler of personalized medicine by drawing attention to any identifiable subgroup of patients defined by the formed strata that may have a notably low likelihood of experiencing longer survival with the test treatment relative to the control treatment despite an overall statistically significant result in favor of the test treatment.

In summary, using a detailed analysis of a hypothetical dataset, retrospective analyses of two real datasets, and results from a simulation study, we have illustrated the impressive power-boosting performance and utility of our proposed 5-step stratified testing and amalgamation routine (5-STAR) relative to that of the logrank test and other common approaches that are not designed to leverage inherently structured patient heterogeneity. We end by making observations on two unrelated but important fronts. First, suppose an interim analysis is planned for a randomized clinical trial to enable a potentially earlier conclusion of either futility or overwhelming success for the test treatment. With application of 5-STAR in mind, it is natural to inquire whether the risk strata formed at the interim analysis should be re-used for the subsequent final analysis (if applicable), or whether the risk strata formation step should be repeated for the latter. Second, even though the focus of this manuscript has been on survival analysis, the 5-STAR concept can be extended to analyses of continuous, ordinal and nominal endpoints. Research on both fronts is ongoing and will be the subject of future communications.

*Software*: the 5-STAR methodology proposed in this manuscript can be easily implemented using the *fiveSTAR* R package available at https://github.com/rmarceauwest/fiveSTAR.